# Non-equilibrium many-particle spin states in self-assembled quantum dot hydrogen, helium and lithium

B. Marquardt,[1] M. Geller,[1, *] B. Baxevanis,[2]
D. Pfannkuche,[2] A. D. Wieck,[3] D. Reuter,[3] and A. Lorke[1]

[1]Fakultät für Physik and CeNIDE, Universität Duisburg-Essen,
Lotharstraße 1, 47048 Duisburg, Germany

[2]I. Institut für Theoretische Physik, Universität Hamburg,
Jungiusstraße 9, 20355 Hamburg, Germany

[3]Lehrstuhl für Angewandte Festkörperphysik, Ruhr-Universität Bochum,
Universitätsstraße 150, 44780 Bochum, Germany

(Dated: July 2, 2010)

**Summary paragraph:**

Epitaxial semiconductor quantum dots (QDs) are nanoscale objects, which incorporate fully confined carriers and are therefore perfectly suited for fundamental studies on atom-like objects in a solid-state environment.[1,2] In addition, these structures are promising building blocks for future quantum information processing.[3] For self-assembled QD, great progress has been made in this field using optical excitation and detection methods.[4–9] Optical techniques, however, may be difficult to integrate with existing (classical) computer technology, and the interpretation of the data is complicated by the fact that different types of carriers, i.e. electrons and holes, are involved. An all-electrical preparation and time-resolved detection of non-equilibrium QD states is therefore highly desirable, but has so far only been achieved in lithographically defined QDs at very low temperatures (mK).[10–13] Here, we introduce an all-electrical measurement technique, which makes it possible to prepare and detect the ground and excited many-particle states in self-assembled InAs quantum dots at liquid helium temperatures. This way, the pure-electron spectra of QD-hydrogen, -helium and -lithium are resolved. Comparison with detailed many-body calculations enables us to identify the different charge and spin configurations in the spectra and in particular detect the singlet and triplet states of QD helium. Furthermore, the time-resolved evolution of the density of states from non-equilibrium to equilibrium charge occupation is shown.

The investigated sample consists of an inverted high electron mobility transistor (HEMT) with an embedded ensemble (> $10^7$) of self-assembled InAs quantum dots (QDs),[1,4,14] see Fig. 1a and supporting material. As the electron channel, which supplies the carriers for the QDs, we employ a two-dimensional electron gas (2DEG).[15] The tunneling barrier between the 2DEG and QDs was chosen so that the tunneling time is in the range of a few milliseconds. When a voltage pulse is applied to the gate, the carrier density and thus the conductivity of the 2DEG will change with a short RC time constant of $T_{RC} = 300$ μs. At the same time, the energy levels of the dots, embedded in the dielectric of the HEMT, will shift[1] and electrons will start to tunnel between the 2DEG and the dots. This will lead to a time-dependent change of the carrier density in the 2DEG,[16] which can be monitored with high resolution by recording its conductivity. Previous studies on similar structures have shown[15] that the change in the conductivity is to a good approximation a linear function of the charge in the dot layer $Q_{QD}$. Applying a constant voltage across source and drain of the HEMT and recording the source-drain current $I_{SD}$ will give a direct measure of the time-dependent tunneling into the dots. Because the response time of the 2DEG to an applied gate voltage pulse is much shorter than the typical tunneling times, the present setup allows us to prepare non-equilibrium situations, where the chemical potential in the 2DEG and in the dot layer differ greatly and tunneling can take place over a wide range of energies (see insets in Fig. 1b). As we will show in the following, this makes it possible to investigate excited QD states, as in $n-i-n$ tunneling structures,[17] however, with adjustable, well-defined initial charge (0 ... 3 electrons per dot).

Figure 1b shows the measurement procedure. First, a well-defined quantum dot state is prepared by setting an initial bias $V_{\text{ini}}$ and giving the system enough time (0.2 s) to equilibrate. Then the probe pulse $V_{\text{p}}$ is applied and $I_{\text{SD}}$ is recorded as a function of time (Fig. 1c). For better signal-to-noise ratio, $I_{\text{SD}}(t)$ is averaged over 1000 pulses. Because, as mentioned above, $\Delta I_{\text{SD}}(t)$ is proportional to $Q_{\text{QD}}$, its derivative with respect to the gate voltage $\frac{d\Delta I_{\text{SD}}}{dV_{\text{p}}}$ is proportional to the (quantum) capacitance of the layer system, which is given by the density of states in the QDs.[15,17] Therefore, by plotting $\frac{d\Delta I_{\text{SD}}}{dV_{\text{p}}}$, the complete time evolution of the density of states can be obtained as the dots are filling up with electrons. Gate voltages $V_{\text{p}}$ can be converted to energies using the well-established lever arm model $\Delta E = e\Delta V_{\text{p}}/\lambda$, where the lever arm $\lambda = \frac{d_{\text{gate}}}{d_{\text{QD}}} = 6$ is given by the distance $d_{\text{QD}}$ between 2DEG and dots and the distance $d_{\text{gate}}$ between 2DEG and gate.

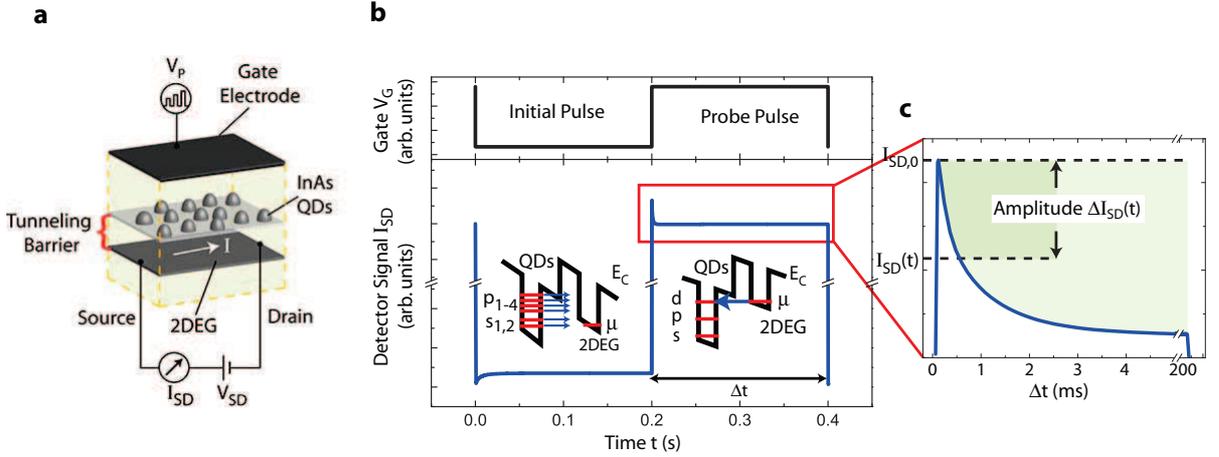

FIG. 1: **Schematic sample geometry and measurement technique.** (a) The transistor structure, which is used to record the time-dependent tunneling into self-assembled quantum dots. A voltage pulse, applied to the gate electrode (b, top), will abruptly change the carrier density in the electron channel (2DEG). The corresponding change in resistance is observed in the time-resolved measurement of the source-drain current $I_{\rm SD}(V_{\rm SD}=$ const.) (b, bottom). The voltage pulse will also strongly shift the chemical potential of the dots, embedded in the dielectric between the gate and the 2DEG. This creates a non-equilibrium situation, where electrons from the 2DEG can tunnel into excited QD states (b, right inset). The time-dependent transfer of charge into the dots, $Q_{\rm QD}(t)$, will cause a decrease in carrier density of the 2DEG, which is detected by recording the change in source-drain current $\Delta I_{\rm SD}(t)$ (c).

Figure 2a displays a three-dimensional plot of such spectra for a transient time $t$ between 0.5 ms and 9 ms. The initial bias is set to $V_{\rm ini} = -1.0$ V, so that the dots are completely empty before $V_{\rm p}$ is applied. The data demonstrates how the density of states in the QD ensemble evolves, as the dots are filled subsequently with up to 6 electrons.

Let us first consider the spectra for the shortest possible time delay $t = 0.5$ ms $\geq T_{RC}$. In the present structure, the tunneling times between the 2DEG and the dots are in the order of $\tau_s = 6$ ms and $\tau_p = 1.4$ ms for the $s$- and $p$-states, respectively,[16] so that at $t = 0.5$ ms, the dots are, to a good approximation, empty. As shown in Fig. 2b, we observe three equidistant peaks in the signal, with a spacing of 55 meV, which is in good agreement with previous studies.[14] However, these measurement shows for the first time degenerate electron states without Coulomb-interaction. The final state configuration of the tunneling process is sketched above each peak: At $V_p = -0.56$ V, electrons resonantly tunnel into the lowest ($s$-) state of the dots; at $V_p = -0.23$ V, non-equilibrium tunneling into the degenerate $p$-states takes place and at $V_p = +0.1$ V, electrons are injected directly into the $d$-shell.

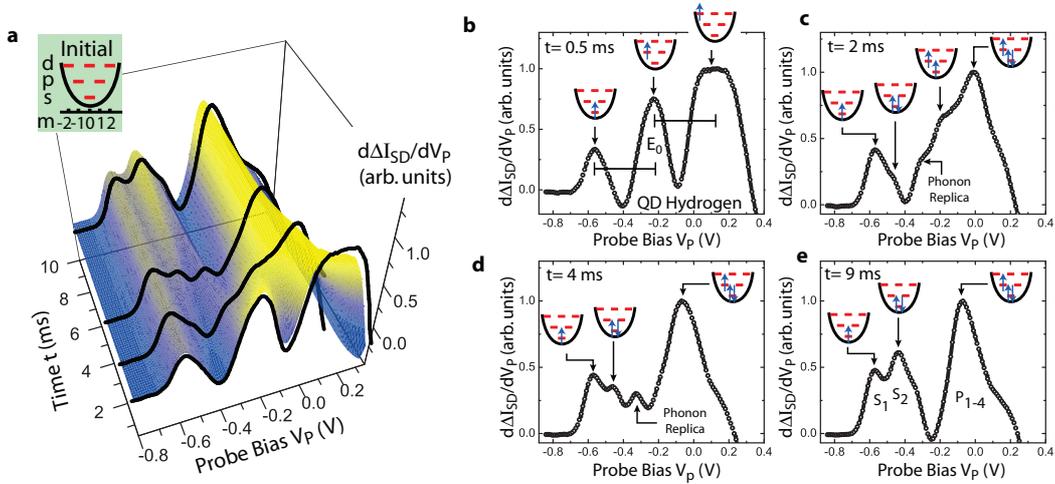

FIG. 2: **Time evolution of the quantum dot charging spectrum.** (a) Coloured surface plot of $d\Delta I_{SD}/dV_p$ – corresponding to the tunneling density of states – as a function of time and pulse bias $V_p$. The initial bias is set so that the QDs are empty at $t = 0$. The inset in (a) illustrates the harmonic model potential, together with the common labels for the QD states and their angular momentum $m$. The diagrams b–e show the charging spectra for different time delays. The electron configurations of the final states, obtained from a comparison between measured and calculated energies, is depicted above the different resonances. For the shortest possible time delay of 0.5 ms (panel b), the data exhibits roughly equidistant charging peaks, as expected from the single particle harmonic oscillator model. Around $t = 2$ ms (panel c), many-particle charging peaks start to appear and tunneling into the ground and excited states of the two-electron system ("QD helium") is visible at $V_p = -0.45$ V and $V_p = -0.2$ V, respectively. An additional structure at $V_p = -0.32$ V (see also panel d) is caused by a phonon-replica of the lowest single-electron charging process. At $t = 9$ ms (panel e), the system is close to equilibrium and the charging peaks resemble those found in capacitance spectroscopy, with two $s$-states ($s_1$ and $s_2$, separated by the Coulomb energy) and a broad structure, comprising the non-resolved resonances of the four $p$ states.

In the opposite limit, $t \gg \tau_s, \tau_p$, the amplitude $\Delta I_{SD}(V_p, t \to \infty)$ corresponds to the total charge in the dots under equilibrium conditions and the signal $\frac{d\Delta I_{SD}}{dV_p}$ will reflect the equilibrium, many-particle spectrum of the dots. For the longest time delay shown here (9 ms), we find two maxima around $V_p = -0.5$ V, and a broad structure ranging from $-0.2$ V to $+0.2$ V. Comparison with results from capacitance spectroscopy[1,16] shows that the first two resonances are the two $s$-states of the QDs, which are separated by $\approx 20$ meV because of Coulomb interaction. The broad structure is a signature of the four resonances of the $p$-shell, which can not be resolved individually.

Figures 2b and c show the spectra for intermediate delay times $t \approx \tau_s, \tau_p$. Features from tunneling into both equilibrium and non-equilibrium states can be observed. For example, the spectrum at $t = 2$ ms exhibits peaks at both $V_p = -0.45$ V and $-0.2$ V. Comparison with theory (see below) allows us to identify these resonances as tunneling into the ground and excited state, respectively, of the doubly occupied dot (QD helium). Also, a new resonance at $-0.32$ V appears, which cannot be explained by an equilibrium or non-equilibrium state in the dots. From the fact that – compared to the lowest ($s_1$) resonance – it exhibits a constant energy difference of about 40 meV and a constant amplitude, we conclude that this resonance is a phonon replica[18] of the $s_1$-state, caused by efficient scattering of tunneling electrons by inelastic LO phonons in the strained GaAs barrier.[19]

Even though many features of equilibrium and non-equilibrium tunneling can be identified in the time-resolved evolution shown in Fig. 2, the fact that signals from dots with different occupations are superimposed makes a detailed interpretation of this data difficult. Therefore, in the following, we will ensure a well-defined tunneling process $n - 1 \to n$ by (1) setting $V_{ini}$ so that $n - 1$ electrons are present in the dots when the probe pulse is applied and (2) considering only the shortest possible time delay, so that higher-order tunneling processes $\Delta n = 2$ are strongly suppressed. The number of electrons in the final state determines which QD 'element' is being investigated: Quantum dot hydrogen, helium or lithium for $n = 1$, 2, 3, respectively.

Quantum dot hydrogen has already been discussed above (see Fig. 2b). Here, the spectroscopy of the equidistant excited energy levels with a spacing of $\hbar\omega = 55$ meV provides valuable input for the theoretical treatment of the many-particle spin states. The spectrum of QD helium is shown in Fig. 3a. We find a resonance at $V_p = -0.45$ V which can be identified as tunneling into the 2-electron ground state ($s_2$), in agreement with equi-librium measurements[16] and the results in Fig. 2. Further resonances can be identified at $V_{\text{p}} = -0.2$ V and $-0.12$ V as well as a broad peak around $V_{\text{p}} = +0.2$ V. Quantum dot lithium is shown in Fig. 3b, where $V_{\text{ini}} = -0.3$ V was chosen so that two electrons occupy the lowest ($s$) shell and the third electron can be injected either into the $p$- or the $d$-shell. Two resonances are observed at $+0.05$ V and $+0.3$ V.

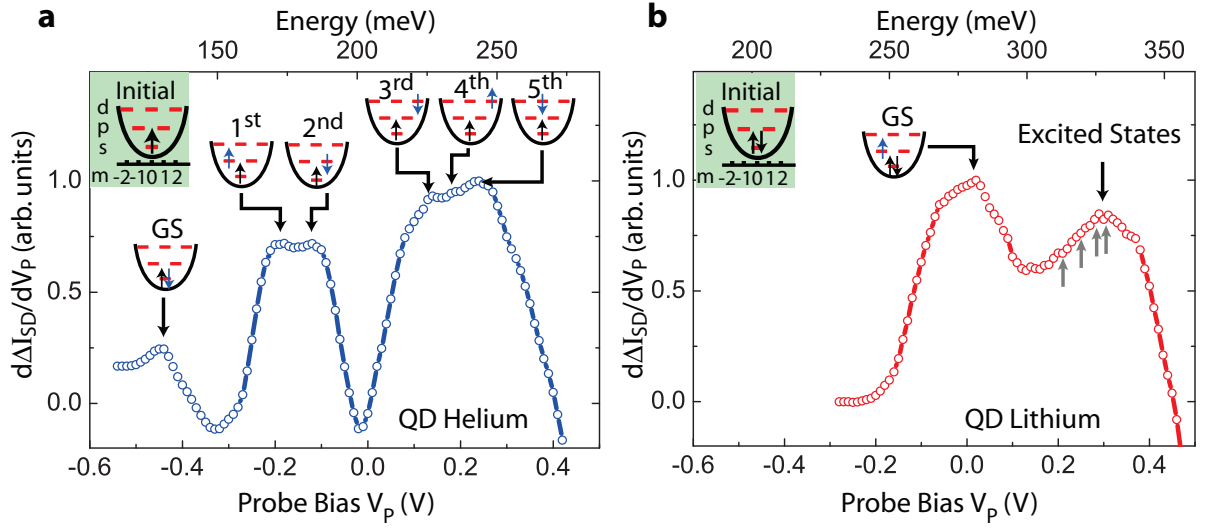

FIG. 3: **Spectroscopy of excited states in QD helium and lithium.** Panels (a) and (b) show the charging spectra for tunneling into the 2- and 3-electron final states, respectively, taken at the shortest possible time delay $t = 0.5$ ms. The initial states, prepared by appropriately setting $V_{\text{ini}}$, are depicted in the upper left corner of each panel. The electron configurations, drawn above the measured resonances, are taken from a comparison with theoretical calculations (see Fig. 4). For QD helium, we observe the lowest resonance at $V_{\text{p}} = -0.45$ V, caused by tunneling into the 2-electron ground state $s_2$ (see also Fig. 2e). Around $-0.15$ V, a double peak structure is seen, corresponding to tunneling into the $p$-shell. The splitting is caused by the difference in exchange energy between the triplet ($V_{\text{p}} = -0.2$ V) and the singlet ($V_{\text{p}} = -0.12$ V) excited state. Three further resonances can be identified around $+0.2$ V, which are caused by tunneling into the excited $d$-shell. For QD lithium, also a clear separation between tunneling into ground and excited states is possible, however, due to the multitude and small separation of the resonances (calculated values are depicted by upward grey arrows), the individual peaks of the excited states cannot be unambiguously identified.

For a thorough identification of the different resonances we have calculated the many-particle energy states in a two-dimensional harmonic oscillator for $n = 1, 2, 3$ electrons using the exact diagonalization method which provides numerically exact solutions.[20,21] This yields the many-particle states of the interacting electrons in terms of superpositions of single-particle Slate determinants. Their coefficients give the probability of single-particle configuration to be found. The level spacing $\hbar\omega = 55$ meV was taken from the single-particle spectrum in Fig. 2b. The effective mass and the dielectric constant were chosen to be $m^* = 0.067m_0$ and $\epsilon_r = 16.5$, respectively, which go into the calculations by the single adjustable parameter $\sqrt{m^*\omega}/\epsilon_r$.[14] In Fig. 4, the calculated energies for the ground state (GS) and the first few excited states are listed, together with the leading terms in the Slater determinant expansion (relative contributions given in %). Also shown are the experimental energies, determined using the resonance condition for tunneling $E^n - E^{n-1} = e\Delta V_{\rm p}/\lambda$ and choosing the zero point of the energy scale such that the single electron ground state corresponds to $E_{GS}^1 = E(s_1) = 55$ meV.

Through a comparison between theory and experiment, we can clearly identify the double peak structure in Fig. 2a ($V_{\rm p} = -0.2$ V and $-0.12$ V) as tunneling into the triplet and singlet spin states of excited QD helium. The energy difference of about 13 meV is a direct measure of the exchange interaction in self-assembled quantum dots, which has so far only been studied in self-assembled dots using optical spectroscopy.[4,8] The splitting is large enough to allow for the all-electrical preparation of spin-polarized states: From a deconvolution of the double peak structure, we estimate that on the low-energy side of the resonance, $V_{\rm p} = -0.26$ V, about 99% of the electrons tunnel into the triplet state.

The higher-lying resonances of QD helium are very closely spaced and therefore hardly discernible in the experiment. However, taking the calculations as a guide, they can be identified as tunneling into different (spin triplet and singlet) $d$-states of the dots as indicated in the insets of Fig. 3. It should be pointed out, though, that two of the corresponding calculated final states have an equally strong contribution from doubly occupied $p$-levels (see Fig. 4, 3rd and 5th excited state). Since the initial state is purely $s$-type, we have neglected these contributions in the assignment of Fig. 3. However, to clarify the role of these contributions in the tunneling process, a full calculation of the tunneling dynamics, taking into account also Auger-type processes would be necessary, which is beyond the scope of this work.

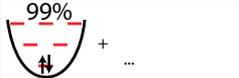

| State # | QD Helium Configuration | $E_{theo}^{He}$ (meV) | $E_{exp}^{He}$ (meV) | State # | QD Lithium Configuration | $E_{theo}^{Li}$ (meV) | $E_{exp}^{Li}$ (meV) |
|---|---|---|---|---|---|---|---|
| GS  | 99% ... | 133 | 130 | GS  | 97% ... | 273 | 281 |
| 1st | 99% ... | 178 | 172 | 1st | 99% ... | 313 |     |
| 2nd | 99% ... | 189 | 184 | 2nd | 66% + 33% ... | 319 |     |
| 3rd | 50% + 50% ... | 231 | 227 | 3rd | 66% + 33% ... | 325 | 328 |
| 4th | 99% ... | 234 | 234 | 4th | 66% + 33% ... | 329 |     |
| 5th | 50% + 50% ... | 240 | 244 |     |     |     |     |

FIG. 4: **Calculated excitation spectra and comparison with the experimental data.** The table lists the energies of the ground state (GS) and $n^{th}$ excited states for QD helium and lithium, calculated using an exact diagonalization method ($E_{theo}$). Also shown are the leading configurations in the Slater determinant expansion with their relative contributions given in percent. For QD helium, the good agreement between $E_{theo}$ and the experimental values $E_{exp}$ makes it possible to identify all peaks observed in the spectra of Fig. 3a. For QD lithium, only the ground state can be determined with certainty. The experimentally determined values have an error of about 3% for QD helium and 5% for QD lithium.

The spectrum for QD lithium consists of two broad maxima. From comparison with theory and near equilibrium measurements[16], we conclude that the low-energy resonance is caused by tunneling into the 3-electron ground state. The additional resonance at $V_p = +0.3$ V (energy separation $\approx 47$ meV) is a clear signature of tunneling into excited QD lithium states. However, because of the multitude and small energy separation of these states (see calculated values, indicated by upward grey arrows), they cannot unambiguously identified.

The measurement technique described here introduces the possibility to study time-resolved, non-equilibrium tunneling phenomena in self-assembled quantum dot systems. This opens up a range of opportunities to investigate the structure and dynamics of excited states in these "artificial atom-like" model systems. The observation of tunneling into both singlet and triplet levels can be considered a first step towards an all-electrical manipulation and detection of spins in epitaxial QDs; – a feat, which has so far only been possible in lithographically defined nanostructures.[12] Our findings also provide a link to the intriguing spin and interaction effects of self-assembled quantum dots, which have been identified using optical spectroscopy.[22] In closing, we would like to mention that the method introduced here for preparation and read-out of quantum states in self-assembled structures has very favorable scaling properties: Because the resistance of a 2DEG is only given by its length-to-width ratio, good signal-to-noise ratio is expected even for scaling down to the single dot level – a prerequisite for quantum information technologies.

Methods

The sample was grown by solid-source molecular beam epitaxy on a GaAs(100) semi-insulating substrate. The active part of the structure starts with 150 nm $Al_{0.34}Ga_{0.66}As$, a Si-$\delta$-doping, a 40 nm $Al_{0.34}Ga_{0.66}As$ spacer with the 2DEG at the edge. What follows is a tunneling barrier of 15 nm GaAs, 10 nm $Al_{0.34}Ga_{0.66}As$ and 5 nm GaAs. On top of the GaAs layer the InAs QDs were grown, followed by another 40 nm GaAs and an insulating AlAs/GaAs superlattice. Finally, the structure was capped with 5 nm GaAs. The dot density of the sample is about $8.3 \times 10^9 cm^{-2}$, determined by atomic force microscopy studies of similarly grown dots on the sample surface. The gated electron channel area is $1.3 \times 10^5 \mu m^2$ which leads to about $1 \times 10^7$ probed QDs. Hall measurements yield a charge carrier density and a mobility of the 2DEG of about $7.4 \times 10^{11} cm^{-2}$ and 9340 $cm^2$/Vs, respectively. All measurements were performed in a He-cryostat at a temperature of 4.2 K, using a pulse generator at the gate contact and a current amplifier with a bandwidth of 1 MHz at the source contact. The drain contact was grounded and a bias of 30 mV was used at the current amplifier to measure the source-drain current $I_{SD}$.